\documentclass{article}

\usepackage{arxiv}
\usepackage{amsmath}
\usepackage[utf8]{inputenc} 
\usepackage[T1]{fontenc}    
\usepackage{hyperref}       
\usepackage{url}            
\usepackage{booktabs}       
\usepackage{amsfonts}       
\usepackage{nicefrac}       
\usepackage{microtype}      
\usepackage{lipsum}
\usepackage{graphicx}
\graphicspath{ {./images/} }

\title{Quantum Hall Resistance and Quantum Hall Plateaus from Edge State Quantization}

\author{
Pedro Pereyra \\
  Departamento de Ciencias B\'asicas, UAM-Azcapotzalco, M\'{e}xico D.F., C. P. 02200, M\'exico \\
  \texttt{ppereyra@azc.uam.mx}
}

\begin{document}
\maketitle
\begin{abstract}
Despite the extensive literature on the quantum Hall effect (QHE), a direct derivation of the phenomenological formula $\rho_{xy} = h/e^2\nu$ from first principles has remained elusive. In this work, we revisit the Landau and Landauer-B\"uttiker formalisms and impose hard-wall boundary conditions on the wavefunction, an essential but often overlooked constraint. This condition quantizes the guiding center position and the longitudinal wave number $k_x$, leading naturally to a discrete number of edge states without invoking energy bending. We derive the Hall resistance directly and recover the standard result $\rho_{xy} = h/e^2\nu$, along with an explicit expression for the filling factor $\nu$ in terms of the Fermi energy and magnetic field. The resulting resistance steps reproduce the observed QHE plateaus and match experimental data without fitting parameters.
\end{abstract}


Since its discovery,\cite{vonKlitzing1980} the quantum Hall effect (QHE) has inspired decades of theoretical and experimental work.\cite{Stormer,Laughlin1983,McIver2020,Yu2010} The integer QHE is conventionally explained using Landau quantization and the emergence of dissipationless edge states, with the Hall resistance given by the phenomenological expression $\rho_{xy} = h/e^2\nu$, where $\nu$ is the filling factor. While several theoretical approaches have reproduced the main features of the QHE, including B\"uttiker's Landauer-based model yielding $\rho_{xy} = h/e^2 N$,\cite{Buttiker1988} a direct derivation of the phenomenological formula from first principles has remained out of reach. Most explanations assume disorder localization, energy levels bending, or phenomenological arguments to reproduce the observed steps in $\rho_{xy}$, but we do not know of direct derivation from wavefunction constraints.

In this work, we revisit the Landau and Landauer approaches with a critical addition: the imposition of hard-wall boundary conditions that force the wavefunction to vanish at the sample edges. This seemingly simple condition leads to the quantization of the longitudinal wave number $k_x$ through a relationship involving Hermite polynomials. As a result, the number of edge channels becomes well-defined and naturally linked to the filling factor $\nu$ and to the Landau level index $n$.

This insight allows us to derive both the conventional Hall resistance and an explicit formula for the filling factor as a function of the Fermi energy and magnetic field. We compare our results with experimental data and find excellent agreement with observed resistance plateaus.

This approach differs from the topological formulations developed by Thouless et al.\cite{Thouless1982} and others, which rely on periodic boundary conditions, Berry curvature, or Chern numbers to explain quantized conductance. While mathematically rigorous derivations such as the one in Ref.\cite{Hastings2015} prove quantization up to exponentially small corrections, they often obscure the physical origin of resistance quantization. In contrast, our derivation remains within standard quantum mechanics, relying solely on the constraint imposed by hard-wall boundary conditions and magnetic confinement. This yields the quantization of longitudinal wave numbers and the number of edge channels in a transparent and physically grounded way.

In the QHE, electrons move confined in a two-dimensional stripe of width $L_y$ in the presence of a perpendicular magnetic field $\mathbf{B} = B\hat{z}$. The Schr\"odinger equation in the Landau gauge $\mathbf{A} = -By\hat{x}$, effective mass and independent particle approximation is:
\begin{equation}\label{SchEq2}
\left[\frac{1}{2m^*} \left(-i\hbar \nabla - e\mathbf{A} \right)^2 + V(y)\right]\Psi(x,y) = E\Psi(x,y),
\end{equation}
where $V(y)$ is an infinite square-well potential: $V(y)=0$ for $|y|<L_y/2$ and $V(y)=\infty$ otherwise. It is well-known that in the absence of the confining potential $V(y)$, a solution of the form $\psi(x,y) = e^{ik_x x}\eta(y)$ transforms the equation into a 1D harmonic oscillator centered at $y_o = -k_x \ell_B^2$, with $\ell_B = \sqrt{\hbar/eB}$ the magnetic length. The resulting equation is:
\begin{equation}
\left[-\frac{d^2}{dy^2} + \frac{1}{\ell_B^4}(y - y_o)^2\right] \eta(y) = \frac{2m^*E}{\hbar^2}\eta(y).
\end{equation}
Its solution is
\begin{equation}
\eta_n(y) = C_n e^{-(y - y_o)^2/2\ell_B^2} H_n\left(\frac{y - y_o}{\ell_B}\right),
\end{equation}
where  $y_o$ is the center of the Landau orbit,  with energy eigenvalues
\begin{equation}
E_n = \hbar\omega_c\left(n + \frac{1}{2}\right), \quad \omega_c = \frac{eB}{m^*}.
\end{equation}
\begin{figure*}[hbt]
\begin{center}
\includegraphics[width=16cm]{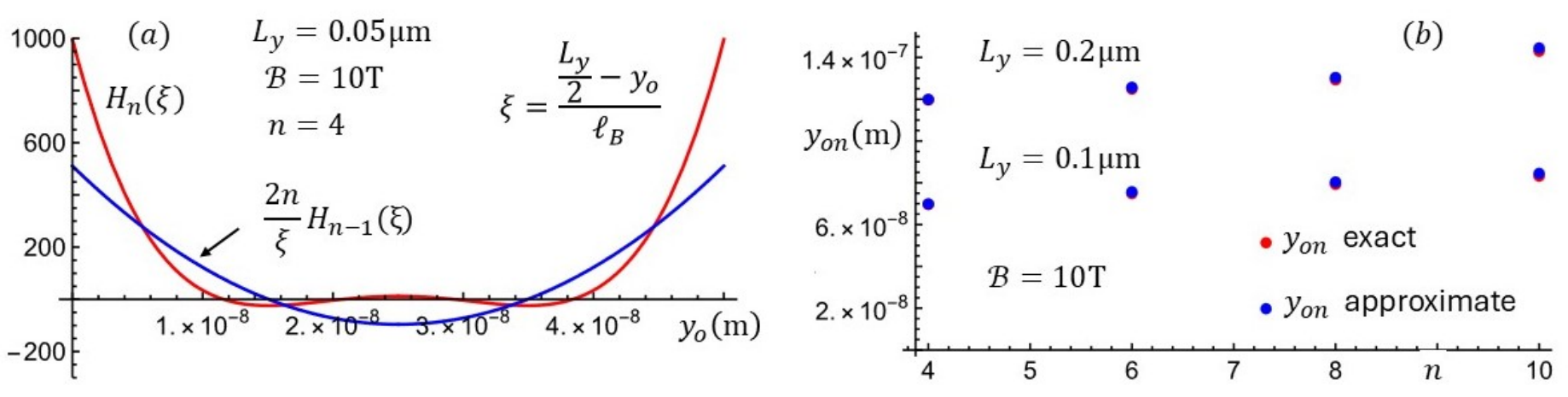}
\caption{In (a) the  guiding center $y_{on}$ at the intersection of $H_n$ and $2n H_{n-1}/\ell_B$, and in (b) the exact values of the guiding center $y_{on}$, closest to the longitudinal edge, predicted by equation (\ref{QuantCond}) (red dots), and the approximate values obtained from the approximate formula (\ref{aproxyon}) (blue dots). For these results the magnetic field is ${\cal B}=$10T and the stripe widths are $L_y=$0.1$\mu$m (lower dots) and $L_y=$0.2$\mu$m (upper dots). } \label{yonQuant}
\end{center}
\end{figure*}

Let us now consider a trial wavefunction:
\begin{equation}
\Psi(x,y) = \sum_j \phi_j(x) \sin\left(\frac{2j\pi y}{L_y}\right) e^{ik_x x} \varphi(y).
\end{equation}
The sine factors ensure that $\Psi$ vanishes at the boundaries $y = \pm L_y/2$. When the energy $E$ is equal  to $E_n$ and $\varphi(y)=\eta_n(y)$, the trial function satisfies the Schrödinger equation if and only if
\begin{equation}\label{EigCond}
  \frac{d\eta_n}{dy}\Bigg|_{y=\pm L_y/2}=0,
\end{equation}
which yields the quantization condition:
\begin{equation}\label{QuantCond}
  H_n\left(\frac{\pm L_y/2-y_o}{l_B}\right)=\frac{2n l_B}{\pm L_y/2-y_o}H_{n-1}\left(\frac{\pm L_y/2-y_o}{l_B}\right),
\end{equation}
This quantization of $y_o$ implies that only a discrete number of states fit within the width $L_y$.

Notice that this condition is  independent of $\phi_j(x)$, which we can choose equal to 1. The accurate values of $y_o$ can be obtained by solving numerically equation (\ref{QuantCond}), but a good approximation to obtain the closest orbit center to the edge is
\begin{equation}\label{aproxyon}
  \frac{L_y/2-y_{on}}{l_B}\simeq-\sqrt{2n-2}.
\end{equation}
\begin{figure*}[hbt]
\begin{center}
\includegraphics[width=11cm]{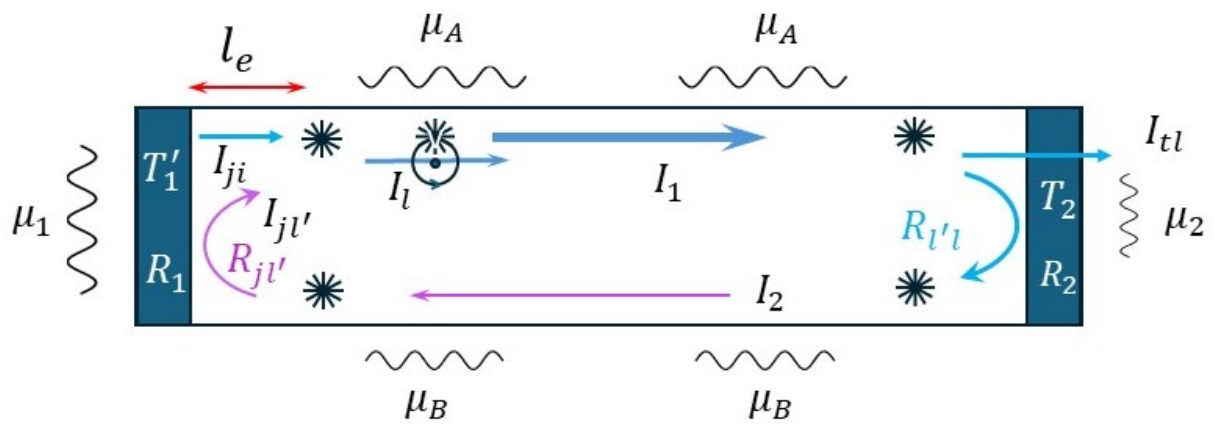}
\caption{The chemical potentials in the source-drain reservoirs and in the probe contacts in equilibrium with the edge states. The transmission and reflection coefficients at the source-drain contacts, refer to the probabilities for the electrons to be transmitted through the contacts into or from the reservoirs or get reflected by the contacts, as explained in the text. The electrons fed into or reflected towards the edge states equilibrate one inelastic length $l_e$ away from the contacts.  In the presence of strong magnetic field, these electrons move in equipotential paths at the upper and lower edges with backscattering suppressed by the magnetic field, and in equilibrium with the non-dissipative probes. } \label{QHallEffect}
\end{center}
\end{figure*}
Given the relation $k_x=-y_o\ell_B^2$, it is clear that the longitudinal wave numbers at the edge of the stripe are also quantized
\begin{equation}\label{aproxkxn}
  k_{xn}=-\frac{y_{on}}{l_B^2}\simeq-\frac{\sqrt{2n-2}}{l_B}- \frac{L_y}{2l_B^2}\hspace{0.2in}{\rm with}\hspace{0.2in}n=1,2,...
\end{equation}
with the same quantum number as the energy. This is an important result that connects the edge states with the Landau energy levels, a relation that will be essential in deriving the Hall resistance in the Landauer approach. In figure \ref{yonQuant} we compare the quantized centroid $y_{on}$ obtained by solving the equation (\ref{QuantCond}) with those obtained with the approximate formula (\ref{aproxyon}). The agreement is good. Taking the quantum number $n$ from the Landau energy $E_n$, and replacing in the approximate formula (\ref{aproxyon}), it is easy to show that
\begin{equation}\label{EnergyEn}
  E_n=\frac{\hbar^2}{2m}\left(k_{xn}+\frac{L_y}{2\ell_B^2} \right)^2+\hbar \omega_c +\frac{1}{2}\hbar \omega_c \hspace{0.2in} n=1,2,...
\end{equation}

This formula shows that the  energy $\hbar \omega_c(n+1/2)$ splits into the kinetic (translational plus rotational) energy, the zero point energy $(1/2)\hbar \omega_c$ and the energy $\hbar \omega_c$ from the zero point energy level to the first energy level. Notice that $k_{xn}$ is defined for $n\geq 1$.
It is important here to emphasize that these energies represent the total energy, and that it is the total energy which is quantized.  The effect of the repulsive potential and the strong magnetic field is to redistribute, at the edge, the quantized energy $E_n$ into the transverse and longitudinal energies.

Using the quantized $k_{xn}$ values and the total energy $E_n$,
we establish that the number of available edge states $\nu$ is:
\begin{equation}\label{lfloor}
\nu = n_\text{max} = \left\lfloor \frac{E_T - \hbar\omega_c/2}{\hbar\omega_c} \right\rfloor.
\end{equation}
where $\lfloor x \rfloor$ means the integer part of $x$. The presence of the \(\tfrac{1}{2} \hbar \omega_c\) zero-point term in Landau levels shifts the onset of the first plateau and controls the spacing between transitions. While often considered a formal quantum correction, this term has a deeper interpretation. In stochastic electrodynamics \cite{Boyer,Marshall1978}, it emerges from equilibrium with a classical background radiation field, providing a semi-classical origin for quantum fluctuations. Though our model is fully quantum mechanical, the sharp cutoff condition in Eq.~(\ref{lfloor}) inherits this structure.\footnote{In stochastic electrodynamics, the zero-point energy is attributed to the particle's coupling with the fluctuating electromagnetic vacuum, whose spectral density is given by $\rho(\omega)=\hbar \omega^3/2\pi^2 c^3$. See \cite{Boyer,Marshall1978}.} in which case $E_T=E_F + \hbar \omega/2$.
\begin{figure*}[hbt]
\begin{center}
\includegraphics[width=15.5cm]{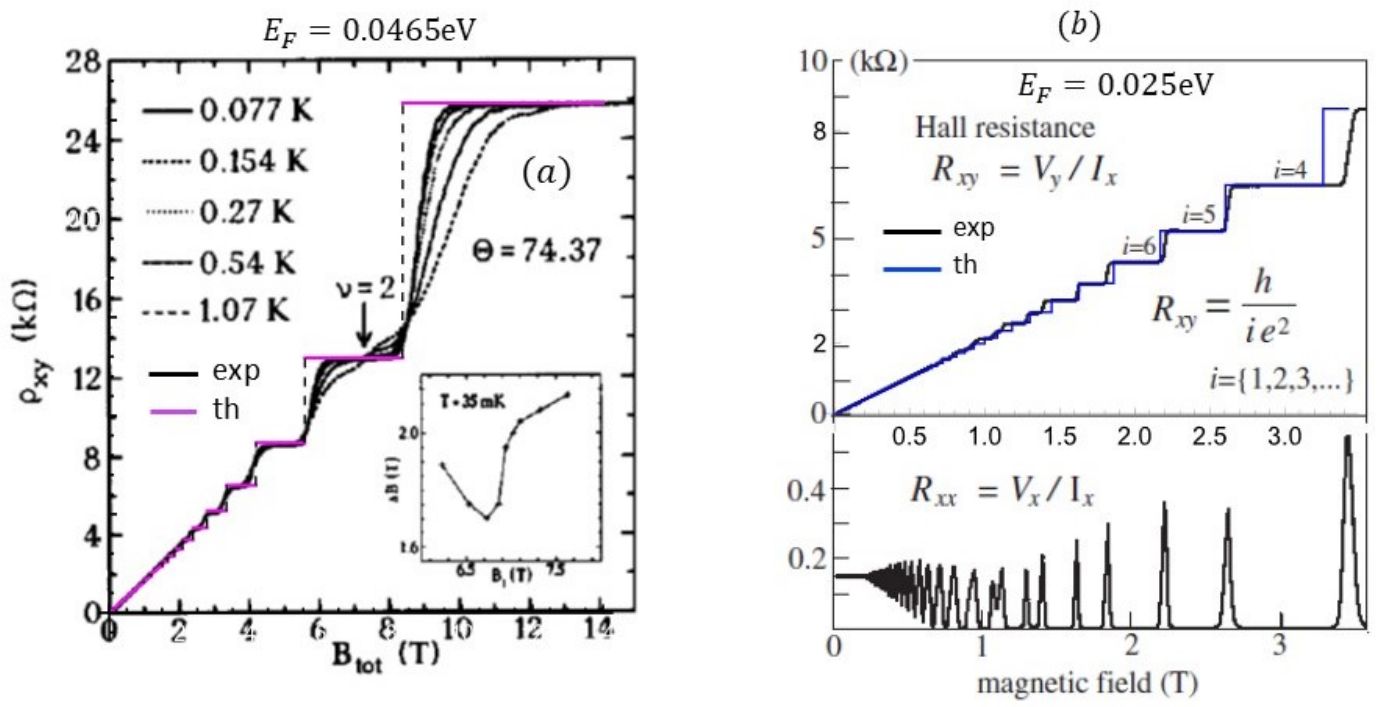}
\caption{Experimental and theoretical Hall resistance (in unit of $h/e^2$). In a) the theoretical prediction compared with the experimental curve measured in a two dimensional electron system at the interface of GaInAs/InP, reported in \cite{Koch1993}. Figure reproduced with the permission of Elsevier. In b) the theoretical prediction compared with the experimental curve measured in a two dimensional electron system embedded in a GaAs/(AlGa)As heterostructure, reported in \cite{WeisVonKlitzing}. Figure reproduced with the permission of The Royal Society (UK). The predicted curves were calculated using the equation (\ref{Hallreslfloor}), for the Fermi energies indicated in the graphs.} \label{QHSsketch}
\end{center}
\end{figure*}
To connect the quantized states with transport properties, we use the Landauer-B\"uttiker approach, sketched in figure \ref{QHallEffect}. Electrons are injected from a reservoir with chemical potential $\mu_1$ and transmitted to another at $\mu_2$ through $\nu$ quantized edge channels. We define $\nu_1$ as the number of channels populated just after the first contact, to distinguish it from the total number $\nu$ that reaches the second contact. We also consider the reflection and transmission coefficients as in Ref. \cite{Buttiker1988}, i.e. $R_1$ and $T_1$
for reflection and transmission through the first contact, for electrons approaching from the right. Thus
\begin{equation}\label{T1}
 T_1=\sum_{j=1}^{M}\sum_{i=1}^{\nu_1}T_{ji}, \hspace{0.2in}R_1=\sum_{i=1}^{\nu_1}\sum_{i'=1}^{\nu}R_{i'i},
\end{equation}
and $R_2$ and $T_2$ for reflection and transmission through the second contact, for electrons approaching from the left. In this case
\begin{equation}\label{T2}
 T_2=\sum_{j=1}^{M}\sum_{i=1}^{\nu}T_{ji}
\hspace{0.2in}{\rm and}\hspace{0.2in}
 R_2=\sum_{i=1}^{\nu}\sum_{i'=1}^{\nu}R_{i'i}.
\end{equation}
Our result aligns with this picture: the discrete jump in \(\nu\) arises from the vanishing of the wavefunction amplitude beyond a cutoff determined by the magnetic confinement, consistent with flux conservation and edge state localization.

Flux conservation implies:
\begin{align}
T_1 + R_1 &= \nu_1, \\
T_2 + R_2 &= \nu.
\end{align}
Even without explicit forms of the transmission and reflection coefficients, it is easy to show, following Büttiker's procedure,\cite{Buttiker1988} that the current $I$ and the chemical potential difference $\mu_A - \mu_B$ can be expressed as:
\begin{equation}
I = \frac{e}{h} \cdot \frac{\nu}{\nu_1\nu - R_1 R_2} T_1 T_2 (\mu_1 - \mu_2),
\end{equation}
and
\begin{equation}\label{Hallvoltage}
  \mu_A-\mu_B=\frac{T_1(\nu-R_2)}{\nu_1 \nu-R_1R_2} (\mu_1-\mu_2).
\end{equation}
Therefore, the Hall voltage $V_H = (\mu_A - \mu_B)/e$ becomes:
\begin{equation}
V_H = \frac{h}{e^2} \cdot \frac{I}{\nu},
\end{equation}
and leads directly to:
\begin{equation}
\rho_{xy} = \frac{h}{e^2\nu}.
\end{equation}

This is precisely the phenomenological formula used to fit the experimental results.
Replacing $\nu$, the Hall resistance becomes\index{QHE!resistance quantization in the Landauer approach}\index{quantum Hall resistance}
\begin{equation}\label{Hallreslfloor}
  \rho_{xy}=\frac{h}{e^2}\frac{1}{\displaystyle{{\lfloor \frac{2\pi m^*_e E_F}{h e{\mathcal{B}}} \rfloor}}}
\end{equation}
This resistance is plotted in figures \ref{QHSsketch} (a) and (b), in units of $h/e^2$ as a function of the magnetic field ${\mathcal{B}}$, and compared with the experimental curves. In (a) the experimental curve measured in a two dimensional electron system at the interface of GaInAs/InP, reported in \cite{Koch1993}, and the theoretical curve assuming an effective mass $m^*_e=0$.041, and Fermi energy $E_F=0.0465$eV. In figure \ref{QHSsketch} (b) the  experimental curve measured in a two dimensional electron system embedded in a GaAs/(AlGa)As heterostructure, reported in \cite{WeisVonKlitzing}, and the theoretical prediction for an effective mass $m^*_e=0$.067, and Fermi energy $E_F=0.025$eV.

We have derived the quantized Hall
resistance directly from wavefunctions constrained by
boundary conditions and magnetic confinement. The
quantization condition for $y_o$ links the number of edge
states to Landau levels and yields a direct formula for
$\rho_{xy}(\mathcal{B}, E_T)$ that matches experiment. This approach clarifies
the physical basis of the QHE. Our result links microscopic wavefunction behavior to macroscopic resistance quantization, reinforcing the conceptual foundation of the quantum Hall effect.


\end{document}